\documentclass[12pt]{iopart}
\usepackage{iopams}
\usepackage{graphicx}

\def\<{\left\langle}
\def\>{\right\rangle}

\begin{document}

\vspace*{-15mm}
\begin{flushright}
MPP-2008-90\\
arXiv:0807.4930\\
\end{flushright}
\vspace*{0.7cm}

\title{Phenomenology of Hybrid Scenarios of Neutrino Dark Energy}

\author{Stefan Antusch$^{1\star}$, Subinoy Das$^{2 \dagger}$ and Koushik Dutta$^{1 \P}$}
\address{$^1$
Max-Planck-Institut f\"ur Physik (Werner-Heisenberg-Institut),
F\"ohringer Ring 6, D-80805 M\"unchen, Germany}
\address{$^2$ Center for Cosmology and Particle Physics, Department of Physics, New York University, New York, NY, 10003, USA} 
\ead{$^\star$antusch@mppmu.mpg.de, $^\dagger$subinoy@nyu.edu, $^\P$koushik@mppmu.mpg.de}

\begin{abstract}
We study the phenomenology of hybrid scenarios of neutrino dark energy, where in addition to a so-called Mass Varying Neutrino (MaVaN) sector a cosmological constant (from a false vacuum) is driving the accelerated expansion of the universe today. For general power law potentials we calculate the effective equation of state parameter $w_{eff}(z)$ in terms of the neutrino mass scale. Due to the interaction of the dark energy field (``acceleron'') with the neutrino sector, $w_{eff}(z)$ is predicted to become smaller than $-1$ for  $z>0$, which could be tested in future cosmological observations.  
For the considered scenarios, the neutrino mass scale additionally determines which fraction of the dark energy is dynamical, and which originates from the ``cosmological constant like'' vacuum energy of the false vacuum.    
On the other hand, the field value of the ``acceleron'' field today as well as the masses of the right-handed neutrinos, which appear in the seesaw-type mechanism for small neutrino masses, are not fixed. This, in principle, allows to realise hybrid scenarios of neutrino dark energy with a ``high-scale'' seesaw where the  right-handed neutrino masses are close to the GUT scale. We also comment on how MaVaN Hybrid Scenarios with  ``high-scale'' seesaw might help to resolve stability problems of dark energy models with non-relativistic neutrinos.  
\end{abstract}

%key words{dark energy, neutrinos}
%Preprint id: JCAP_040P_1007

%\pacs{}
%\submitto{}

\section{Introduction}
The evidence for the existence of a dark sector in the universe has made the present era of cosmology fascinating and challenging. At present, the microscopic natures of dark energy and dark matter are still an open question, with both components only probed gravitationally \cite{Colless:2001gk,Riess:1998cb,Perlmutter:1998np,Spergel:2006hy}. Regarding dark matter, there are various particle physics candidates which may belong to the class of Weakly Interacting Massive Particles (WIMPS) or which may interact only gravitationally. On the other hand, for particle physics explanations of the observed dark energy, the main fundamental question is whether dark energy is a cosmological constant or a dynamical field. While in the former case the equation of state parameter $w$ of dark energy is constant and equal to $-1$, in the latter case it is a function of redshift $z$ and in general differs from $-1$, providing a way to distinguish both fundamental types of dark energy experimentally. Such a deviation of $w$ from $-1$ will be searched for in various future surveys, studying supernovae, baryonic acoustic oscillations, weak gravitational lensing effects, galaxy clusters, or other techniques \cite{Albrecht:2006um}.
One obstacle to relate dark energy more explicitly to particle physics is the fact that the amount of the observed energy density $\rho_{DE}^{1/4} \sim 0.003$ eV is {\em much} smaller than the ``theoretical expectation''. In this paper we will not address this ``cosmological constant'' problem, i.e.\ the suppression/cancellation of the various generic contributions to dark energy which are too large, but assume that it is resolved by some other  mechanism.  

Our work is motivated by the intriguing observation that the energy density of dark energy is close to another very small scale in particle physics, namely the one of neutrino masses. The discovery of flavour conversion of neutrinos from various sources, interpreted within the framework of neutrino oscillations, points to two mass eigenvalues of the light neutrinos above about $0.01$ eV and $0.05$ eV \cite{Mohapatra:2005wg}, while searches for neutrino masses from Tritium $\beta$-decay and neutrinoless double $\beta$-decay yield an upper bound for each mass eigenvalue of roughly $0.5$ eV \cite{Mohapatra:2005wg}.  
The proximity of the scale of dark energy to that of neutrino masses has inspired the proposal of the so-called Mass Varying Neutrino (MaVaN) scenario \cite{Fardon:2003eh}. In this scenario, the relic neutrino density contributes to the effective potential for the dynamical field which is called ``acceleron''. One consequence is that neutrino masses vary on cosmological time scales and, since the relation between neutrino masses and $\rho_{DE}^{1/4}$ remains valid in earlier cosmic epochs, also dark energy behaves differently from a cosmological constant.  
Compared to other dynamical models of dark energy (e.g.\ quintessence \cite{Wetterich:1987fm,Peebles:1987ek,Caldwell:1997ii}), in the MaVaN scenario one does not have to choose a mass for the dynamical field of the order of the present Hubble parameter $H_0$. In fact it can be much larger, even of the order of the neutrino mass scale, due to the stabilising effect of the contribution of the 
relic neutrino density to the potential. As a result, the field adiabatically tracks the minimum of the effective potential whose evolution is controlled by the time evolution of neutrino number density.

The MaVaN models have been investigated in many studies, e.g.\ regarding possible experimental signatures and constraints \cite{Kaplan:2004dq,Zurek:2004vd,Li:2004tq,Barger:2005mn,Schwetz:2005fy,Ringwald:2006ks, Cirelli:2005sg, Weiner:2005ac, Brookfield:2005bz,Bernardini:2008pn} as well as model-building issues \cite{Fardon:2005wc, Takahashi:2005kw, Ma:2006mr, Takahashi:2007ru, Bjaelde:2008yd}. 
It has turned out that while there are many interesting possible signatures and attractive features, the  scenario is tightly constrained by the requirement of consistency with late time structure formation. In particular, it has been pointed out \cite{Afshordi:2005ym} that if neutrinos are non-relativistic today, due to fifth force effects the neutrinos would cluster at late time and finally form ``neutrino nuggets'' which would spoil the dark energy behavior of the neutrino fluid. 

The goal of this paper is to investigate the phenomenology of a generalised scenario which we will refer to as MaVaN Hybrid Scenario, where in addition to a MaVaN sector a cosmological constant (from a false vacuum) is driving the accelerated expansion of the universe today.
As we will show, for a generalised power law potential in the MaVaN sector the effective equation of state parameter $w_{eff}(z)$ as well as the fraction to which the dark energy is of dynamical nature is determined by the neutrino mass scale. 
We will also analyse the possibility to realise neutrino dark energy with a ``high-scale'' seesaw mechanism, where the right-handed neutrino masses are close to the GUT scale, and comment on how such models might help to suppress the formation of ``neutrino nuggets'' and resolve stability problems of MaVaN models with non-relativistic neutrinos. For the main part of the paper we will focus on neutrinos that are non-relativistic today.

The paper is organised as follows: In Section 2 we will review the basics of the MaVaN scenario. Section 3 contains the definition of the phenomenological framework and the analysis is performed in Section 4. In Section 5 we discuss the possibility of realising neutrino dark energy with a ``high-scale'' seesaw mechanism and in Section 6 we comment on how this might help to solve stability issues for non-relativistic neutrinos. Section 7 contains our Summary and Conclusions.

\section{The Original MaVaN Scenario}\label{Sec:Scenarios}
 
In the basic MaVaN scenario a singlet fermion $N$ (right-handed neutrino) and a dynamical real scalar field $A$ (called  ``acceleron'') are introduced. Their dynamics gives rise to the acceleration of the universe and to  neutrino masses which vary with time. The basic form of the Lagrangian is given by
\begin{equation}
{\cal L} \supset m_{D} \nu N + \frac{1}{2} \kappa A N N + \mbox{H.c.} + V(A)\;,
\end{equation}
where $m_D$ is neutrino Dirac mass, $\kappa$ is a dimensionless coupling in dark sector and $V(A)$ is the potential energy of the acceleron field. 
The Standard Model neutrino $\nu$ as well as the singlet neutrino $N$ are both written as two-component left-chiral Weyl spinor fields. 
If $\kappa A \gg m_D$, i.e.\ if the Majorana mass of the right-handed neutrino is much larger than the neutrino Dirac mass, one can effectively integrate out $N$ from the low energy theory which leads to the effective Lagrangian 
\begin{equation}
{\cal L} \supset -\frac{1}{2} \frac{m_{D}^2}{\kappa A} \nu \nu  + \mbox{H.c.} + V(A)\;.
\end{equation}
The neutrino mass $m_\nu = |\frac{m_{D}^2}{\kappa A}|$ depends on the dynamics of $A$.
Taking into account the masses of the non-relativistic relic neutrinos we now include a term proportional to the relic neutrino number density $n_{\nu}$ to the effective potential, 
\begin{equation}
V_{eff} = n_{\nu} \left\vert\frac{m_{D}^2}{\kappa A}\right\vert + V(A) \;.
\end{equation}

The effective potential $V_{eff}$ for the acceleron $A$ has a minimum
because the first term pushes $A$ to larger values whereas as the
second term $V(A)$ is assumed to favour smaller values of $A$ (e.g.\
if $V(A) \propto \log(A/\Lambda), {\rm or} \propto A^{\alpha}$).
Typically, in this type of models a sub-eV scale is 
introduced in order to realize the desired values of $V(A)$ (and finally of $V_{eff}(A)$).\footnote{
A scenario of quintessence cosmology with growing matter component (which may be applied, for instance, 
to neutrinos) has recently been proposed without explicitly invoking any sub-eV scale 
\cite{Amendola:2007yx,Wetterich:2007kr}.
}
As a consequence of the two contributions to the potential and since $n_\nu$ decreases with time, the field value of the acceleron field $A$ becomes smaller with time when it adiabatically tracks the minima of the potential (if the mass of $A$ is above the Hubble parameter). The combined $A$ and neutrino fluid drives the accelerated expansion of the universe. 

We remark that unlike in conventional quintessence models the scalar field does not roll down an extremely flat potential but it rather gets stabilised by the finite neutrino density. For this reason, the slope of the potential can be much higher than in quintessence scenarios and the mass of the scalar field $A$ can even be as high as the neutrino mass scale, whereas in quintessence models it has to be as tiny as the Hubble scale.  

Although the model can successfully explain the present acceleration of the universe with masses of the dynamical scalar field much larger than the Hubble scale (e.g.\ in the sub-eV range) and provide a connection between dark energy and the physics of neutrino masses, it has also some shortcomings: For example, the potential is not stable under radiative corrections, unless new states appear with sub-eV masses, and it still has to be rather flat in order to be consistent with present constraints on the equation of state parameter. 
Furthermore, the MaVaN scenario with non-relativistic neutrinos has a problem with respect to the consistency with late time structure formation, as has been pointed out in Ref.~\cite{Afshordi:2005ym}. If neutrinos are non-relativistic today, due to fifth force effects the neutrinos could cluster at late time and finally form ``neutrino nuggets'' which would spoil the dark energy behavior of the neutrino fluid. 

However, it has been demonstrated that under certain conditions these
problems might be overcome \cite{Fardon:2005wc}. 
For example, in supersymmetric theories of neutrino dark energy the potential can be stable under radiative corrections (for example in gauge mediated supersymmetry breaking where supersymmetry breaking is only transmitted to the dark sector via gravity such that the dark sector superpartners are light) and stability problems with respect to neutrino density perturbations can be avoided if neutrinos are highly relativistic,  although then the connection to neutrino masses is weakened since the two observed mass scales correspond to non-relativistic neutrinos.

\section{Generalised Framework: The MaVaN Hybrid Scenario}\label{Sec:MaVaNHybrid}

After having reviewed the basics of the original MaVaN scenario in the last section, we will now set up the framework for our analysis. It consists of a generalised MaVaN potential of power law type plus a constant vacuum energy contribution. In the following we will specify it explicitly and comment on how it may be realised in classes of models as well as on the connection to the (in principle) experimentally measurable neutrino mass scale.

\subsection{Generalised Potential for Neutrino Dark Energy}
In the following we will consider a rather general setup where
neutrino dark energy is realised in a type of ``hybrid dark energy''
setup with the theory being in a false vacuum at present time. This
implies that in addition to the time varying contribution of the
acceleron field there is an additional contribution of the energy
density of the false vacuum.\footnote{One such example model is the
  SUSY version of the MaVaN scenario proposed in \cite{Fardon:2005wc}.}
Motivated by the notion of providing a rather general framework for our study, we will therefore consider an effective potential for neutrino dark energy of the form 
\begin{equation}\label{effective_potential}
V(A)_{eff} = V(A) + \frac{\rho_{\nu}^{(0)}}{a^3} \left(\frac{A_0}{ A}\right)^{\beta} +  V_{0} , 
\end{equation}
where 
\begin{equation}\label{potential}
V(A) = M^{4- \alpha}A^{\alpha}.
\end{equation}
$A$ is the acceleron field\footnote{For simplicity we will take $A$ as a real scalar field as in Section \ref{Sec:Scenarios}, unless stated otherwise.}, $\rho^{(0)}_\nu$ is the energy density of the SM neutrinos today and $a$ is the scale factor.
The main ingredients are a constant term $V_0$ (from a false vacuum), a potential for $A$ which is monotonically increasing with $|A|$ (in our case a general power law potential $\propto A^\alpha$) and an effective term $n_\nu m_\nu$, where we assume $m_\nu \propto 1/A^\beta$ in order to generalise the conventional seesaw relation. We note that the potential of Eq.~(\ref{potential}) is not the most general form of a potential that may arise in a realistic model. Instead of the monomial potential that we are considering, it can for instance also be a polynomial or logarithmic function of the field $A$, or a combination of both.

\subsection{Possible Origin of the Generalised Potential}
To motivate the appearance of the constant vacuum energy in the ``hybrid scenario'' by an example, we may consider the following superpotential (after EW symmetry breaking),  
\begin{eqnarray}\label{Eq:Origin_1}
W = \lambda \hat A (\hat N^2 - v_N^2) + y \hat H^0_u \hat \nu  \hat N + \frac{1}{2} m^2_A \hat A^2\;,
\end{eqnarray} 
where hats indicate superfields.
$\hat N$ denotes the SM singlet (right-handed neutrino) superfield(s), and the most relevant contributions to the scalar potential stem from the F-terms 
$|F_{A}|^2 + |F_{N}|^2= |\lambda v_N^2|^2 + |2 \lambda A N + m_A A + ...|^2$. 
The first term contributes the constant vacuum energy while the second term gives a mass term ($\sim |A|^2$) to the acceleron field. Together with the additional effective potential term $n_\nu m_\nu \sim n_\nu v_u^2 / |A|$, where $v_u$ is the vev of the Higgs field which couples to the neutrinos, the potential for the acceleron has a minimum for a non-zero field value of $A$. $|F_N|^2$ also gives a time-varying mass to the right-handed neutrino, $m_N \sim |A|$. 

While the above superpotential may serve as a simple example which could realise the generalised potential of Eq.~(\ref{effective_potential}) with $\alpha=2$ and $\beta=1$, it should be kept in mind that that there is in principle a large variety of possible models which can give rise to other values of $\alpha$ and $\beta$ (as well as to other non-power-law forms of the potential). 
For example, different powers of $A$ may appear in the superpotential or the neutrino masses may originate from a different type of seesaw mechanism (e.g.\ from a so-called double \cite{Bhatt:2007ah} or multiple seesaw) such that $n_\nu m_\nu \sim n_\nu v_u^2 / |A|^\beta$, and so on. 
Furthermore, one may anticipate that the parameters $\alpha$ and $\beta$ in realistic values should be positive integers. However, it is also possible that fractional values of the exponents appear, for example if the kinetic terms in a model are non-canonical and therefore a field transformation has to be performed in order to  bring the kinetic terms back to canonical form. We will discuss the possible origin of the generalised potential in more detail in the Appendix.

\subsection{Connection to Neutrino Masses}\label{Sec:ConnectionToMnu}
From neutrino oscillation experiments, two so-called mass squared differences are known: $\Delta m^2_{31} = m_3^2 - m_1^2$ is known mainly from atmospheric neutrinos and $\Delta m^2_{21} = m_2^2 - m_1^2$ from solar neutrino data and from the KamLAND experiment. The experimental values are about $\Delta m^2_{21}=7.6 \times 10^{-5} \:\mbox{eV}^2$, $|\Delta m^2_{31}| = 2.5 \times 10^{-3} \:\mbox{eV}^2$ \cite{Mohapatra:2005wg}. With this nomenclature we also know that $m_2 > m_1$, but $m_3$ can be the heaviest or the lightest of the mass eigenvalues. For the sum $\bar m_\nu$ defined as 
\begin{eqnarray}
\bar m_\nu = \sum_{i=1}^{3} m_i
\end{eqnarray}
we know that there is a lower bound from oscillations (saturated if $m_1 \approx 0$ and $m_2 \approx 0.01$ eV, $m_3 \approx 0.05$ eV)  and also an upper bound from neutrinoless double $\beta$ experiments of about $1.5$ eV (with $m_1 \approx m_2 \approx m_3 \approx 0.5$ eV) \cite{Mohapatra:2005wg}. Finally, we have approximately
\begin{eqnarray}\label{Eq:mbar_range}
\bar m_\nu \in [0.06 \:\mbox{eV}, 1.5 \:\mbox{eV}] \;.
\end{eqnarray}
On the other hand, the lightest of the neutrino mass eigenvalues ($m_1$ or $m_3$) can be arbitrarily small.

One important question is whether this sum $\bar m_\nu$ is relevant for the MaVaN setup, or whether the mass of only one of the light neutrinos matters. To answer this question, we have to go to the full three family scenario where we distinguish two possible cases.

In one possible situation (which we may call case I) $\bar m_\nu$ is indeed relevant. This case applies when the vev of the field $A$ gives masses to all the right-handed neutrinos $N_i$ such that the part of the Lagrangian relevant for neutrino masses reads for instance ($X_{ij}$ is a coupling matrix)
\begin{eqnarray}
{\cal L} \supset \frac{1}{2} X_{ij} A N_i N_j + (Y_\nu)_{\alpha i} \nu_\alpha H_u^0 N_i \; .
\end{eqnarray} 
The seesaw suppressed mass matrix of the light neutrinos is then given by
\begin{eqnarray}
(m_\nu)_{\alpha\beta} = \frac{v^2\,(Y_\nu)_{\alpha i} X^{-1}_{ij} (Y_\nu)^T_{j \beta}}{A} \;.
\end{eqnarray}
If we diagonalise $m_\nu$, the effective potential reads 
\begin{eqnarray}
V_{eff} = \sum_\alpha n_{\nu_\alpha} (\bar m_{\nu}^{\mathrm{diag}})_{\alpha\alpha} = n_\nu \sum_\alpha (\bar m_{\nu}^{\mathrm{diag}})_{\alpha\alpha} = n_\nu \bar m_\nu \;,
\end{eqnarray}
where we have assumed that $n_{\nu_\alpha} = n_\nu$ for all generations $\alpha$ of neutrinos. 

On the other hand, we could have the situation (which we may call case II) that each $N_i$ gets its mass from a different scalar field, e.g.\ from the vevs of fields $A,B$ and $C$ and that we have a Lagrangian of the form
\begin{eqnarray}
{\cal L} &\supset&  \frac{1}{2} x_1 A N_1 N_1 +  \frac{1}{2} x_2 B N_2 N_2 +  \frac{1}{2} x_3 C N_3 N_3 \nonumber \\
&& + \sum_i (y_\nu)_{i} \nu_i H_u^0 N_i \; ,
\end{eqnarray} 
where $x_{i}$ are coupling constants.
Now the light neutrino mass matrix would be given by 
\begin{eqnarray}
(m_\nu)_{\alpha\beta} = 
\frac{v^2\,(y_\nu)_{1}^2}{ x_1 A} 
+\frac{v^2\,(y_\nu)_{2}^2 }{ x_2 B} 
+\frac{v^2\,(y_\nu)_{3}^2}{ x_3 C} \;.
\end{eqnarray}
In this example, only $m_1$ would be relevant for the effective potential for $A$.\footnote{Of course, analogously, one can realise the case where any of the mass eigenvalues of the light neutrinos is relevant for the effective potential for $A$.} If consistency of the MaVaN scenario (for a particular form of the potential) would require a very light neutrino mass scale $\ll 0.01$ eV, this could be realised in case II. 
The relevant mass would then be $m_1$ for a so-called normal hierarchical spectrum or $m_3$ for an inverted hierarchical spectrum.
 
In the following, we will mainly be interested in case I, since it has a particularly close link to the observed neutrino masses which manifests itself in the upper and lower experimental bounds on today's value of $\bar m_\nu$ given in Eq.~(\ref{Eq:mbar_range}). We will therefore present our results using the notation for case I where the relevant neutrino mass scale is $\bar m_\nu$. Most of the formulae can readily be adopted to case II, as long as neutrinos remain non-relativistic.

\section{Phenomenology of the MaVaN Hybrid Scenario}
In this section we will investigate the dynamics and phenomenology of the MaVaN Hybrid Scenario with a general power law potential 
\begin{eqnarray}
V(A)= M^{4-\alpha}A^{\alpha}\;.
\end{eqnarray} 
The observed small neutrino masses are generated by a version of the seesaw mechanism, and we will assume a general dependence of the neutrino masses on $A$ of the form 
\begin{eqnarray}
\bar m_{\nu}(A) \sim \frac{v^2}{A^\beta}\;,
\end{eqnarray} 
where $v$ is the vev of the SM Higgs field (which is proportional to the Dirac masses of the neutrinos). The notation $\bar m_{\nu}$ is introduced in section \ref{Sec:ConnectionToMnu}.

\subsection{Background Evolution}
As a first step of our analysis we consider the background dynamics of the scalar field $A$ in the MaVaN Hybrid Scenario defined in section \ref{Sec:MaVaNHybrid}, which leads to an effective potential for $A$ of the form 
\begin{equation} \label{effectiveVback}
V(A)_{eff} =  M^{4- \alpha}A^{\alpha} + \frac{\rho_{\nu}^{(0)}}{a^3} \frac{\bar m_{\nu}(A)}{\bar m_\nu^{(0)}}  +  V_{0}\;, 
\end{equation}
where $n_{\nu}^{(0)} \bar m_{\nu}^{(0)} =\rho_{\nu}^{(0)}$ 
and where the index ``$(0)$'' indicates todays values of the parameters. 
The second term in the potential originates from the coupling between neutrino and the scalar field in the effective Lagrangian. 
In the following discussions, if we do not mention the values of $\alpha$ and $\beta$, we will assume $\alpha =2$ and $\beta =1$ which corresponds to the ``standard'' MaVaN scenario, i.e.\ where the potential for $A$ is quadratic and where the seesaw mechanism is of type I. 
We will also focus on the case where the neutrinos are non-relativistic today. 

The Friedmann equation in our scenario takes the form
\begin{equation}
3 H^2 M_{Pl}^2 = \frac{\rho_{CDM}^{(0)}}{a^3} + \frac{\rho^{(0)}_{Baryons}}{a^3} + V(A) + \frac{\rho^{(0)}_\nu}{a^3} \left(\frac{A_0}{ A}\right)^\beta+ \frac{1}{2}\dot A^2  + V_{0} \label{Friedmann1}
\end{equation}
and the scalar field $A$ obeys the following modified Klein Gordon equation
\begin{equation}
\ddot{A} + 3 H \dot{A}= -\frac{dV(A)}{dA} - \frac{\rho_{\nu}^{(0)}}{a^3} \frac{\frac{d \bar m_{\nu}}{dA}}{\bar m_{\nu}^{(0)}}\;. \label{KG}
\end{equation}
We note that while the dark matter density $\rho_{CDM}$ redshifts as $a^{-3}$, the time dependence of the neutrino energy density is non-trivial due to the time-dependence of $A$.

Now we use {\it adiabaticity}, i.e.\ the feature of MaVaN scenario that the acceleron field adiabatically follows the minimum of the effective potential since long time back in the cosmic history. Consequently, the condition $\frac{\partial V_{eff}}{\partial A} = 0$ determines the evolution of the scalar field to be
\begin{equation}\label{Eq:AinMinimum}
A = \left(  \frac{\beta\rho_{\nu}^{(0)}A_{0}^{\beta}}{a^3 \alpha} M^{-(4-\alpha)}  \right)^{\frac{1}{\alpha +\beta}}. \label{A}
\end{equation}
The present value of $A_0$ can be found from the above equation and is given by 
\begin{equation}
A_{0} = \left(\frac{\beta\rho_{\nu}^{(0)}}{\alpha} M^{-(4-\alpha)}    \right)^{1/\alpha}\;, \label{AoverAnot1}
\end{equation}
which allows to derive the relation
\begin{equation}
\frac{A}{A_0} = \left(1+z\right)^{\frac{3}{ \alpha +\beta}}\;, \label{sol1}
\end{equation}
where $a=1/(1+z)$ has been used with $z$ being the redshift.
From this expression we can calculate the kinetic energy of the field compared to the total energy density carried by the field,
\begin{equation}
\frac{\dot A ^2/2}{\rho_A} = 1.5 \left(\frac{A_0}{M_{Pl}}\right)^2 \left(\frac{1}{\alpha +\beta}\right)^2 (1+z)^{\frac{6}{\alpha + \beta}}\:. \label{kinoverdensity}
\end{equation}
We can see that if the field value of $A$ today is smaller than the Planck scale $M_{Pl}$, the kinetic energy of the field can consistently be neglected for the relevant time of the evolution of the dark energy\footnote{On the other hand we will see that a high value of $A_0 \lesssim M_{Pl}$ is typically required with respect to the stability issues for non-relativistic neutrinos (c.f.\ discussion in section \ref{stability}).}. In addition, a  constant vacuum energy contribution further reduces the significance of the kinetic energy compared to the Hubble constant during the evolution. We will discuss the adiabaticity condition of the acceleron field in Section \ref{The MaVaN Hybrid Scenarios with High Scale Seesaw Mechanism}.

Using Eq.~(\ref{sol1}) and comparing the potential energy $V(A)$ to the acceleron-dependent neutrino fluid density in Eq.~(\ref{effectiveVback}), we find that the ratio of the cosmological density parameters of these two components is given by
\begin{equation}
\frac{\Omega_{\nu}}{\Omega_{A}} = \frac{\alpha}{\beta}\;, 
\label{ratio}
\end{equation}
where we have substituted $ 3 H^2 \Omega_{\nu}=\frac{\rho_{\nu}^{(0)}}{a^3}(\frac{A_0}{ A})^{\beta}$ and $3H^2 \Omega_{A}= V(A)$. We would like to highlight the result that the ratio of the energy densities from the interaction term and from the potential term $V(A)$ is constant in time. This feature, which is basically a by-product of the adiabaticity condition, will be useful to derive constraints on the model parameters from cosmological observations and it will lead to a particularly close connection to the neutrino mass scale in the MaVaN Hybrid scenario with power law potentials.

\subsection{Dark Energy Effective Equation of State in the General MaVaN Hybrid Scenario}

Using the above results for the background evolution, we will now investigate how the parameters $\alpha$, $\beta$ and $V_0$ of the model are constrained by cosmological observations, in particularly regarding the dark energy equation of state, as well as by the present bounds on (and eventually by a future measurement of) the  neutrino mass scale $\bar m_\nu$.\footnote{ 
We would like to note that only the terrestrial bound on neutrino masses can be applied directly to MaVaN models. Cosmological bounds on neutrino masses are generically relaxed for mass varying neutrinos (see e.g.\ \cite{Ichiki:2008rh}) since the neutrino masses decrease with increasing $z$, as we will discuss below. 
In fact, one ``smoking gun'' signal of MaVaN scenarios would be a ``cosmological upper bound'' on the sum of neutrino masses (if they are assumed as constant in time) which is incompatible with todays lower bound from terrestrial experiments.}

We will start by deriving the effective dark energy equation of state parameter $w_{eff}(z)$. The main point here is that when the dark energy equation of state is extracted from cosmological data sensitive to the Hubble scale $H(z)$, it is always assumed that the dark matter component redshifts as $1/a^3$. 
Massive neutrinos also contribute a small fraction to the dark matter today. The total dark matter density today is then given by 
\begin{eqnarray}
\rho_{DM}^{(0)} = \rho_{CDM}^{(0)} + \rho_{DM}^{\nu (0)}\;,
\end{eqnarray}
where $\rho_{CDM}^{(0)}$ is the amount of cold dark matter and where $\rho_{DM}^{\nu (0)}$ is calculated from the actual value of the neutrino mass $\bar m_\nu^{(0)}$ (which may be determined more precisely by future experiments), i.e.\ $\rho_{DM}^{\nu (0)} = \rho_{\nu}^{(0)}$.
However, as we have already mentioned, the neutrino masses vary with time and therefore this contribution does not redshift as $1/a^3$. Consequently, an observer will not extract the intrinsic dark energy equation of state, but rather an effective one, $w_{eff}$.    

More explicitly, the effective dark energy density (which defines the effective equation of state $w_{eff}$) is extracted from the Hubble equation (assuming for simplicity a flat universe)
\begin{eqnarray}\label{Eq:weffFromHubble}
3H^2 M_{Pl}^2 
= \frac{\rho_{CDM}^{(0)}}{a^3}+\frac{\rho_{Baryons}^{(0)}}{a^3} + \frac{\rho_{DM}^{\nu (0)}}{a^3} + \rho_{DE}^{eff} \;,\label{app}
\end{eqnarray}
where it is (wrongly, but for a model-independent analysis unavoidably) assumed that the neutrino density redshifts like ordinary matter.

We would like to add a few remarks to clarify the notation:
$\rho_{DM}^{\nu (0)}$ (i.e.\ the value of the quantity
$\rho_{DM}^{\nu}$ today) is same as $\rho_{\nu}^{(0)}$. However, in an
earlier epoch, the true neutrino energy density $\rho_{\nu}$ is {\it not} equal to the quantity $\rho_{DM}^{\nu} = \rho_{DM}^{\nu (0)}/a^3$.

To calculate $\rho_{DE}^{eff}$, following \cite{Das:2005yj}, we rewrite Eq.~(\ref{Friedmann1}) as 
\begin{eqnarray}
3 H^2 M_{Pl}^2 &=& \frac{\rho_{CDM}^{(0)}}{a^3} +\frac{\rho_{Baryons}^{(0)}}{a^3}
+\frac{\rho_{DM}^{(0)\nu}}{a^3} -\frac{\rho_{DM}^{(0)\nu}}{a^3} \nonumber \\
&&
+ \frac{\rho_{\nu}^{(0)}}{a^3} \frac{\bar m_{\nu}(A)}{\bar m_\nu^{(0)}} \break + V(A) + \frac{1}{2} \dot {A}^2 + V_{0}\;.
\end{eqnarray}
Comparing this relation with Eq.~(\ref{app}) yields
\begin{equation}
\rho_{DE}^{eff}= \left( \frac{\rho_{\nu}^{(0)}}{a^3}  \frac{\bar m_{\nu}(A)}{\bar m_\nu^{(0)}} - \frac{\rho_{DM}^{(0)\nu}}{a^3}\right)  + V(A) +\frac{1}{2} \dot{A}^{2} + V_{0}\;. \label{dark_energy_eff}
\end{equation}

From $\rho_{DE}^{eff}$ the observed equation of state $w_{eff}$ can be calculated via the continuity equation 
\begin{equation}\label{Eq:Defweff}
\frac{d\rho_{DE}^{eff}}{dt} = -3H(1+w_{eff})\rho_{DE}^{eff}\;. \label{weffective}
\end{equation}
Taking the time derivative of Eq.~(\ref{dark_energy_eff}) 
and using Eq.~(\ref{KG}) and the expression for $\dot{\rho_{A}}$, we find
\begin{equation}
\frac{d\rho_{DE}^{eff}}{dt} = -3H\frac{\rho_{\nu}^{(0)}}{a^3}\left(\frac{\bar m_{\nu}(A)}{\bar m_{\nu}^{(0)}}- 1\right)-3H\dot{A}^2 \:.
\end{equation}
We compare the above equation with Eq.~(\ref{weffective}) to obtain 
\begin{equation}\label{Eq:weffgeneral}
w_{eff} 
=- \frac{V(A) +V_{0} - \frac{{\dot A}^2}{2}  }{\frac{\rho_{\nu}^{(0)}}{a^3} \left( \frac{\bar m_{\nu}(A)}{\bar m_{\nu}^{(0)}} -1 \right) + V(A)+ V_{0} + \frac{{\dot A}^2}{2} }\;. \label{effective_equation}
\end {equation}
This is the general result for the effective equation of state parameter $w_{eff}(z)$ for general $A$-dependent neutrino masses $\bar m_{\nu}(A)$ and for a general acceleron potential $V(A)$. 

We now discuss some of the characteristic feature of $w_{eff}$: 
To start with, it can be seen from the above expression that at present time ($z=0$), the $\rho_{\nu}^{(0)}$-dependent term disappears from the denominator. Furthermore, in the absence of the constant vacuum energy term $V_0$, the equation of state parameter at $z=0$ is simply given by the conventional equation of state parameter for the scalar field $A$,
\begin{equation}
w_{A} = \frac{ \dot{A}^2/2 -V(A)}{ \dot{A}^2/2 + V(A)}\:.
\end{equation} 
Since the slow-roll approximation can be applied for field values less
than $M_{Pl}$ and since the kinetic energy $\sim \dot{A}^2$ is much smaller
than $V(A)$ in this case, $w_A$ is very close to -1. 
Regarding the effective equation of state parameter $w_{eff}$ for $z=0$, adding the constant vacuum energy term $V_0$ even makes its value closer to $-1$.      
  
For $z>0$, however, the neutrino density term in the denominator of Eq.~(\ref{Eq:weffgeneral}) is non-zero. 
Furthermore, noting that in the MaVaN-type models of interest $\bar m_{\nu}(A) < \bar m_{\nu}^{(0)}$,  the $\rho_{\nu}$-dependent term is negative which implies that $w_{eff}$ becomes less than $-1$ for $z>0$.

In summary, the effective equation of state parameter $w_{eff}(z)$ is defined by Eq.~(\ref{Eq:Defweff}), with $\rho_{DE}^{eff}$ related to the Hubble parameter $H(z)$ by Eq.~(\ref{Eq:weffFromHubble}). 
By definition, $w_{eff}(z)$ contains information about whether the effective dark energy density is dynamical or behaves like a cosmological constant at a given redshift $z$. In order to extract $\rho_{DE}^{eff}$ from $H(z)$, additional information regarding the matter components has to be used.   
For example, SN data provides some information on $H(z)$. However, $\rho_{CDM}^{(0)} + \rho_{Baryons}^{(0)}$ is required as an input, in addition to the neutrino energy density $\rho_{DM}^{(0)\nu}$ which depends on the (in principle) measurable neutrino mass scale today. One possibility to obtain $\rho_{CDM}^{(0)}$ and $\rho_{Baryons}^{(0)}$ is to extract it from the CMB, where the neutrinos were relativistic and do not contribute to the matter density. On the other hand, if information on the matter density from other observations, e.g.\ from weak lensing, is used it may already contain $\rho^{\nu}$ if the neutrinos are clustered on the relevant scales.         
We note that in order to confront the MaVaN Hybrid Scenarios with the whole variety of (future) data from cosmological observations, such issues have to be taken into account carefully. 
Finally, in order to test a particular dark energy model of this type using all available cosmological data, one would have to go beyond the analysis of $w_{eff}(z)$ and fit all model parameters to the data.

\subsection{Dark Energy Effective Equation of State in the MaVaN Hybrid Scenario with Power Law Potentials}
We will now quantify the above general statements about $w_{eff}(z)$ for the more explicit scenario where the MaVaN potential has power law form, and investigate the further phenomenological consequences. 
We start with rewriting the general form of the effective equation of state in terms of the parameters of the MaVaN Hybrid Scenario, which (at this stage) are $\alpha, \beta,$ and $V_0$ (assuming further a fixed/measured value of $\bar m_\nu$).
To do so, we use the results from our adiabatic background solution 
$\rho_{\nu}= \frac{\alpha}{\beta} \rho_A$ (see Eq.~\ref{ratio}) and the relation 
\begin{equation}
 \rho_{\nu} (z) = \rho_{\nu}^{(0)} (1+z)^{\frac{3 \alpha}{\alpha + \beta}} \; \mbox{or}\quad 
 \bar m_{\nu} (z) = \bar m_\nu^{(0)} (1+z)^{-\frac{3 \beta}{\alpha + \beta}}
 \;,
 \end{equation}
which is plotted in Fig.~(\ref{mnu_z}) for illustration. 

As the neutrino mass scale increases with time, neutrinos were relativistic 
at some earlier epoch. For example, for $\frac{\alpha}{\beta} = 2$, this
happens around redshift $z = 100$. We note that at this (or larger) redshift, the above-derived
formulae do no longer apply since we have assumed non-relativistic neutrinos 
throughout. 
The field $A$ remains stabilised when the neutrinos are relativistic\footnote{See e.g.\ \cite{Peccei:2004sz} for a discussion in the original MaVaN scenario which can readily be generalised to the hybrid case.},  
however at values which in general differ from those given in Eq.~(\ref{Eq:AinMinimum}).  
During CMB formation time where $z \sim 1000$, neutrinos were highly relativistic 
and furthermore their mass was much smaller than today, which implies that the effects of their mass 
density can be neglected during this epoch.
Details about cosmological bounds from CMB data and large scale structure observations (including 
neutrino masses) can be found, e.g., in \cite{Brookfield:2005bz,Ichiki:2008rh}.

Using further the constraint from the observed dark energy density today,
\begin{equation}
\rho_{DE}^{(0)}= V_0 + \rho_A^{(0)}\;,  \label{constrain}
\end{equation} 
we find that  
\begin{equation}
\frac{\rho_{\nu}^{(0)}}{\rho_{DE}^{(0)}}= \frac{\alpha}{\beta}(1- \gamma), \label{ratio1}
\end{equation}
where we have defined the parameter 
\begin{eqnarray}
\gamma = \frac{V_0}{\rho_{DE}^{(0)}}
\end{eqnarray} 
for the fractional contribution of $V_0$ to the total observed dark energy density today. After some 
straightforward calculations, we finally get the expression for the equation of state in terms of the 
parameters $\alpha, \beta, \gamma $ and as a function of redshift $z$,
\begin{equation}\label{Eq:weffGamma}
w_{eff}(z)=- \frac{\gamma + (1 - \gamma)(1+z)^{\frac{3\alpha}{\alpha + \beta}}}     {\gamma + (1 + \frac{\alpha}{\beta})(1-\gamma)(1 + z)^{\frac{3\alpha}{\alpha + \beta}} - \frac{\alpha}{\beta}(1- \gamma)(1+z)^3}\:. \label{finaleos}
\end{equation}
We can see from the above expression that for $\gamma =1$, $w_{eff}(z)=-1$ as this should be the case for a constant energy density. 
However for $\gamma \neq 1$ we find that $w_{eff}(z)$ deviates from $-1$ for $z>0$. 
The amount of deviation depends on the choice of parameters $\alpha, \beta, \gamma$. 

We can also see that the value of the $\bar m_\nu^{(0)}$ does not appear in Eq.~(\ref{Eq:weffGamma}). As we will show in the next subsection, $\bar m_\nu^{(0)}$ is determined in this setup by the parameters $\alpha, \beta$ and $\gamma$. In fact, we use this to trade the parameter $\gamma$ for the measurable quantity $\bar m_\nu^{(0)}$ and re-express $w_{eff}(z)$ as a function of $\bar m_\nu^{(0)}$.

\subsection{The Effective Dark Energy Equation of State and the Neutrino Mass Scale} \label{predictionforneutrino}

As mentioned above, the neutrino mass scale $\bar m_\nu^{(0)}$ can be calculated as a function of the model parameters $\alpha, \beta$ and $\gamma$. For non-relativistic neutrinos (corresponding roughly to $m_i \gtrsim 10^{-4}$ eV for $i=1,2,3$) the neutrino energy density is $\rho_{\nu}^{(0)} = \bar m_{\nu}^{(0)}  n_{\nu}^{(0)}$. As we have mentioned in section \ref{Sec:ConnectionToMnu}, we are considering case I where the acceleron field provides the masses of all three right-handed neutrinos. Furthermore, the observed dark energy density today is given by 
$\rho_{DE}^{(0)} = V_{0} + V(A_{0})$.
Using these two equations one can show that $\bar m_{\nu}^{(0)}$ can be expressed in terms of $\gamma$ as 
\begin{equation} 
\bar m_{\nu}^{(0)}=(1-\gamma)\frac{\alpha}{\beta}\frac{\rho_{DE}^{(0)}}{n_{\nu}^{(0)}}\;.
\end{equation}
Plugging in the observed value $\rho_{DE}^{(0)} \sim 4 \times 10^{-11} \:\mbox{eV}^4$ and the (standard) theoretical prediction for the neutrino number density today, $n_{\nu}^{(0)} \sim 8.8 \times 10^{-13} \:\mbox{eV}^3$ \cite{Takahashi:2005kw, Kolb:1988aj}, we find 
\begin{equation}\label{Eq:MnuvsGamma}
\bar m_{\nu}^{(0)} \approx 45(1-\gamma)\frac{\alpha}{\beta} \:\mbox{eV}\;. \label{m_nu}
\end{equation}  
This simple expression relates the present value of the neutrino mass to the quantity $\gamma$ which specifies which fraction of dark energy is constant vacuum energy. For fixed $\gamma$, the neutrino mass scale $\bar m_{\nu}^{(0)}$ is predicted in this scenario. 

\begin{figure}
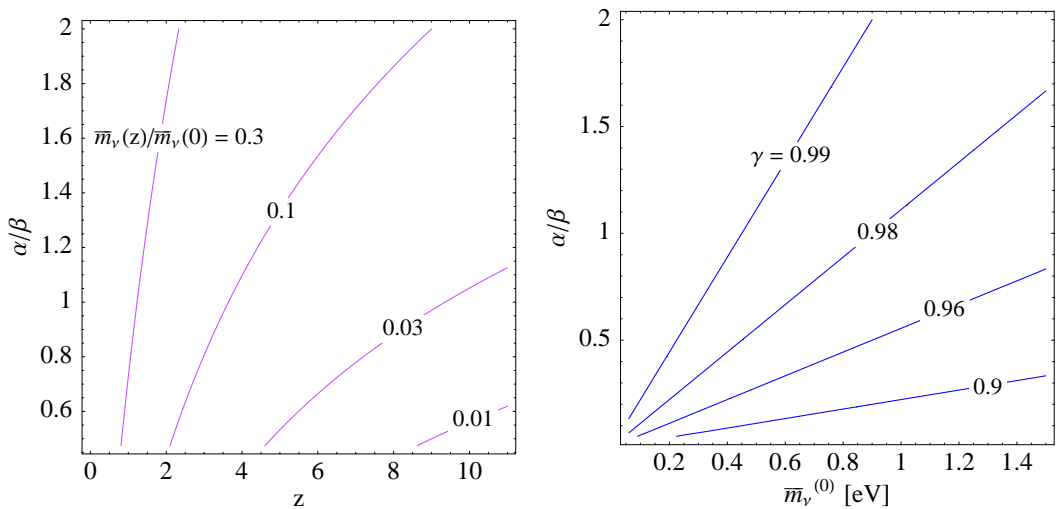

\centering
\includegraphics[scale=0.8]{mnuovermnu0_z_aoverb.eps}\quad
\includegraphics[scale=0.8]{gamma_mnu_aoverb.eps}
\caption{The left panel shows the neutrino mass scale $\bar m_\nu = \sum_{i=1}^3 m_{i}$, divided by their present value, as a function of redshift and of $\alpha/\beta$. 
The right panel shows the value of $\gamma = V_0 / \rho_{DE}^{(0)}$ as a function of 
the neutrino mass scale today and of $\alpha/\beta$. }\label{mnu_z}\label{neutrino_constraint}
\end{figure}

One immediate consequence of the above relation is that for $\gamma = 0$, i.e.\ without a constant energy density contribution $V_{0}$, the predicted neutrino mass violates the the present experimental bounds, unless either $\alpha$ is very small or $\beta$ is very large. 
In particular, for the standard case $\alpha=2$ and $\beta=1$, and for $\bar m_{\nu}^{(0)}$ of the order  
$1$ eV, $V_{0}$ must contribute $\sim 99$\% of todays dark energy  of the total observed dark energy density (c.f.\ Fig. \ref{neutrino_constraint}).
On the other hand, for smaller $\alpha / \beta$ the value of $\gamma$ decreases and a larger fraction of dark energy is dynamical. For example, for $\beta = 1, \alpha = 0.1$ and $\bar m_{\nu}^{(0)} \sim 1$ eV, the dynamical component amounts about $22$\% of the total dark energy density. 

\begin{figure}[t]
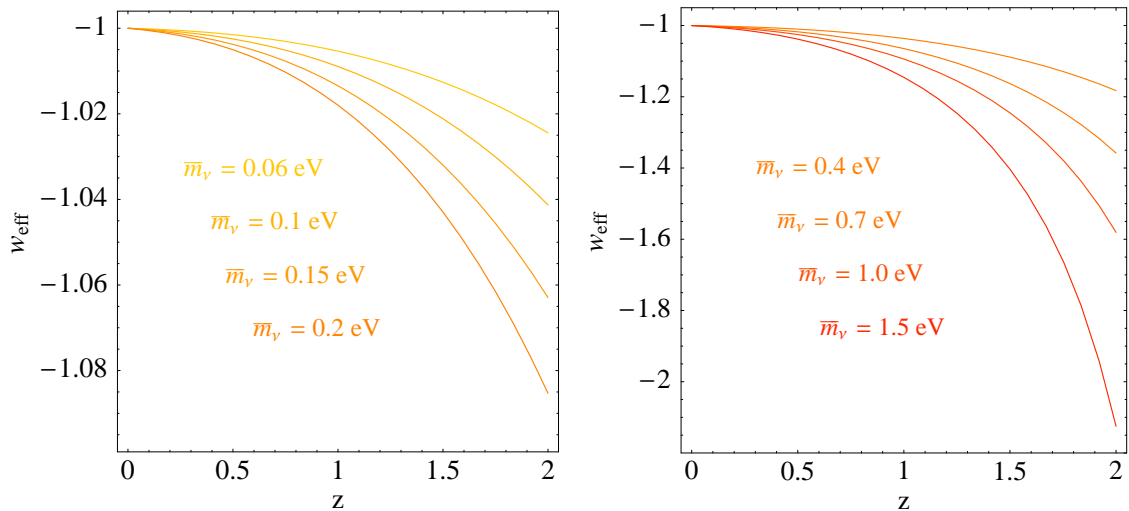

\center
\includegraphics[scale=0.875]{weff_mnu_z_plot1.eps}\quad
\includegraphics[scale=0.857]{weff_mnu_z_plot2.eps}
\caption{\label{weffective_plot} The plots show the effective dark energy equation of state parameter $w_{eff}$ as a function of redshift for different values of the neutrino mass scale $\bar m_\nu = \sum_{i=1}^3 m_{i}$. The plots show the case $\alpha =2, \beta =1$ and the left panel includes examples with $\bar m_\nu^{(0)} \le 0.2$ eV whereas in the right panel values $\bar m_\nu^{(0)} \ge 0.4$ eV are presented.}
\end{figure}

Using Eq.~(\ref{m_nu}) we can express the effective equation of state parameter of Eq.~(\ref{finaleos}) in terms of $\bar m_{\nu}^{(0)}$ by eliminating the parameter $\gamma$, which yields
\begin{equation}\label{Eq:weff_mnu}
\!\!\!\!\!\!\!\!\!\!\!\!\!\!\!\!\!\!
w_{eff} = -\frac  {\left(1 - \frac{\beta}{\alpha}\left(\frac{\bar m_{\nu}^{(0)}}{45~eV}\right) \right ) +  \frac{\beta}{\alpha}(\frac{\bar m_{\nu}^{(0)}}{45~eV})    (1+z)^{\frac{3\alpha}{\alpha + \beta}}   }       {\left(1 - \frac{\beta}{\alpha} (\frac{\bar m_{\nu}^{(0)}}{45 ~eV})  \right)  + \left(1 + \frac{\beta}{\alpha}\right)    (\frac{\bar m_{\nu}^{(0)}}{45~ eV})    (1+z)^{\frac{3\alpha}{\alpha + \beta}}  - (\frac{\bar m_{\nu}^{(0)}}{45 ~eV})    (1+z)^3  }\;.
\end{equation}
In Fig. \ref{weffective_plot} the $z$-dependent effective equation of state parameter $w_{eff}$ is plotted for the case $\alpha = 2$, $\beta =1$ and various values of $\bar m_{\nu}^{(0)}$.

\subsection{Comparison with Present Observations: Averaged Dark Energy Equation of State}
\begin{figure}[t]
\center
\includegraphics[scale=0.875]{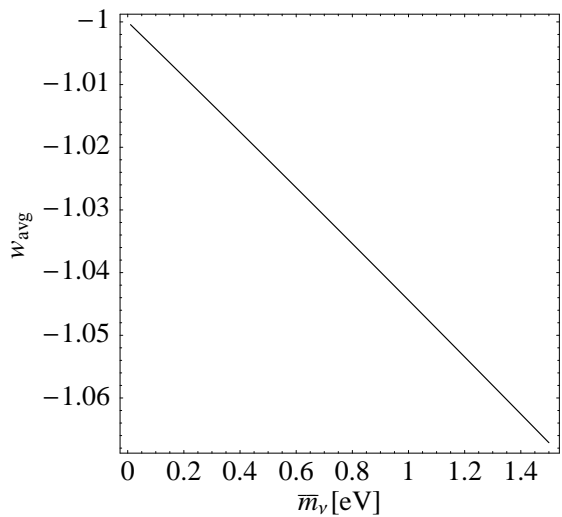}
\caption{\label{w_av}
Averaged equation of state parameter $w_{avg}$ as a function of neutrino mass scale $\bar m_\nu = \sum_{i=1}^3 m_{i}$. 
The plot show the case where $\alpha =2$ and $\beta =1$.}
\end{figure}

For a dark energy equation of state which varies with time, a model independent extraction of 
$w_{eff}(z)$ is difficult from the currently available data. Therefore, in most analyses of the present data, a constant $w$ is assumed. One quantity which is used in the literature to compare with the best fit result under the assumption of a constant $w$ is the weighted (or averaged) equation of state parameter $w_{avg}$ defined in the following way \cite{Das:2005yj}
\begin{equation}
w_{avg} = \frac{\int_{0}^{z} w_{eff}(z)\Omega_{DE}^{eff}(z) ~dz }  {\int_{0}^{z} \Omega_{DE}^{eff}(z) ~dz}.
\end{equation}
In Fig.~\ref{w_av}, $w_{avg} $ is plotted as a function of $\bar m_{\nu}^{(0)}$ for the case $\alpha =2, \beta =1$. 
The WMAP 5 year  $2\sigma$ limit on the constant dark energy equation of state parameter is $-0.097 < 1+w < 0.142$ for zero curvature and $-0.11 < 1+w < 0.14$ when a nonzero value of the curvature is allowed \cite{Komatsu:2008hk}. From the plot we can see that even $\bar m_{\nu}^{(0)}$ as large as $1.5$ eV is compatible with the present data.

\section{The MaVaN Hybrid Scenarios with High Scale Seesaw Mechanism} \label{The MaVaN Hybrid Scenarios with High Scale Seesaw Mechanism}

One interesting result of the previous section is that with power law MaVaN potentials as in Eq.~(\ref{effective_potential}), the predictions for the observables do not depend on the value of $A$ today. 
With the masses $M_R$ of the right-handed neutrinos originating from terms in the potential of the form
\begin{eqnarray}
{\mathcal L }_{\mathrm{M_R}} = \frac{1}{2} \lambda A N N
\end{eqnarray}
this means that in principle the masses of the right-handed neutrinos today ($\sim \lambda A_0$) 
could be close to the GUT scale and we might realise the MaVaN Hybrid Scenario with a ``high-scale'' seesaw mechanism. Compared to the usual MaVaN scenario where the right-handed neutrino masses are very small and where tiny neutrino Yukawa couplings have to be postulated, in the ``high-scale'' seesaw the neutrino Yukawa couplings can be of ${\cal O}(1)$ and the smallness of the neutrino masses is explained by the large masses of the right-handed neutrinos.

Recalling Eq.~(\ref{AoverAnot1}), the scales $A_0$ and $M$ in the considered scenarios are connected to the neutrino mass scale $\bar m_{\nu}^{(0)}$ by 
\begin{equation}\label{scales}
A_0 M^{\frac{4-\alpha}{\alpha}} = \left(\frac{\beta}{\alpha} \bar m_{\nu}^{(0)} n_{\nu}^{(0)}\right)^{1/\alpha}.
\end{equation}
To check the adiabaticity condition for the case of large $A_0$, we consider the square of the effective mass of the acceleron field today which is given by 
\begin{equation}\label{meffsq}
m_{eff}^2 \equiv \frac{d^2 V_{eff}}{d A^2} =  \alpha (\alpha + \beta)\left (\beta / \alpha\right)^{1 - \frac{2}{\alpha}} {\rho_{\nu}^{(0)}}^{1- \frac{2}{\alpha}}      M^{\frac {2}{\alpha} (4-\alpha)}\;.
\end{equation}
Using Eq.~(\ref{scales}) in Eq.~(\ref{meffsq}) we can eliminate the dependence on $M$ and obtain 
\begin{equation}
m_{eff} = \frac{1}{A_0}(\beta(\alpha +\beta)\, \bar m_{\nu}^{(0)} n_{\nu}^{(0)})^{1/2}.
\end{equation}
Now, for adiabaticity to hold, $m_{eff}$ must be larger than the present value of the Hubble scale $\sim 1.5 \times 10^{-33} eV$. Assuming for example $\alpha = 2$ and $\beta =1$ (i.e.\ a simple mass term potential for $A$ and the standard seesaw relation), with $A_0$ around the GUT scale $\sim 10^{16}$ GeV and $\bar m_{\nu}^{(0)}= 1.5$ eV, $m_{eff}$ is about $2 \times 10^{-31}$ eV, which is considerably heavier than the Hubble scale. For the same choice of parameters and present vev of acceleron field, but with $\bar m_{\nu}^{(0)}= 0.06$ eV; $m_{eff} \sim 4 \times 10^{-32}$ eV, still larger than the Hubble scale. 
This is in contrast to quintessence-type models, where the mass of the relevant scalar field has to be below the present Hubble scale. 

We would however like to note at this point that although the scenario has several attractive features, we have not attempted the construction of a viable model of this type. One challenge is the radiative stability of the potential. Another challenge is the origin of the constant vacuum energy contribution. Both issues might be addressed in supersymmetric models (for example in simple setups of the form of Eq.~(\ref{Eq:Origin_1})), however even if hidden sector supersymmetry breaking is mediated to the SM fields by gauge interactions and to the dark sector only by gravity, this would typically induce a too large mass for $A$ of at least about $10^{-4}$ eV. Leaving these model-building questions aside, we will now comment on how the ``high-scale'' seesaw might help with respect to stability issues for non-relativistic neutrinos.

\section{High Scale Seesaw and Stability Issues for Non-relativistic Neutrinos}  \label{stability}
In this section we will comment on the stability issue regarding the neutrino density perturbations,  
which are typically considered as a serious problem 
for MaVaN scenarios with non-relativistic neutrinos. Let us first briefly review the problem: 
Soon after the introduction of the original MaVaN scenario \cite{Fardon:2003eh}, it has been
pointed out \cite{ Afshordi:2005ym} that the original scenario faces a catastrophic 
instability for non-relativistic neutrinos due to the extra force
carried between neutrinos and the acceleron field $A$.
With the mass $m_A$ of the field $A$ much larger than the Hubble scale,
it has been argued that the adiabatic perturbations at scales 
between $m_A^{-1} < \lambda < H^{-1}$ are unstable.

Let us now discuss how this problem might be overcome in the MaVaN Hybrid Scenarios with a 
``high-scale'' seesaw mechanism. 
Rather than attempting a full numerical analysis of the density 
perturbations of the neutrinos in
this paper (which we will leave for a future study), we will provide qualitative 
arguments and analytical estimates which will show how the MaVaN Hybrid Scenarios might help 
to suppress dangerous instabilities.  
We will start with a general consideration regarding the coupling between the neutrinos and the
acceleron field, and then turn to analytical criteria for the occurrence of the instabilities. 

One main difference between the MaVaN Hybrid Scenarios with power law type  potentials (where a 
``high-scale'' seesaw mechanism is in principle possible) and the standard MaVaN scenario is 
that in the former the present value $A_0$ of the acceleron field can
be much larger than the eV scale, even close to the GUT or
Planck scale. 
One can readily see that this might help with respect to the stability issue, when 
the heavy right-handed neutrino fields $N$ are integrated out the theory. 
Then, the interaction term between the light neutrino(s) and the acceleron field becomes
(for, e.g., $\beta = 1$)
\begin{equation}\label{Eq:stab_1}
{\mathcal L }_{\mathrm{int}} =
 \bar \nu_L \nu_L\frac{y^2v^2}{A},
\end{equation}
where $A = A_{0} + \delta A$ with $\delta A$ denoting the quantum fluctuations
around the classical value of the field $A_0$. 
With $\delta A \ll  A_0$, the term in Eq.~(\ref{Eq:stab_1}) takes the form  
\begin{eqnarray}
{\mathcal L }_{\mathrm{int}} \approx  \bar \nu_L \nu_L 
\frac{\bar m_{\nu}^{(0)}}{A_0}\,\delta A\;,
\end{eqnarray} 
where $\bar m_{\nu}^{(0)} =
\frac{y^2v^2}{A_{0}}$. In the usual MaVaN scenario the field value $A_0$ is not much larger than $\bar m_{\nu}^{(0)}$. In contrast to this, as we have discussed in the previous section, 
in the MaVaN Hybrid Scenarios with power law potentials the
present value of $A_0$ can in principle be as large as the GUT scale (or Planck scale)
and the coupling between neutrinos and $A$ can be strongly suppressed (by a factor $\bar m_{\nu}^{(0)}/A_0$).

Let us now turn to the analytical criteria:
First of all we note that when revisiting the stability issues of the MaVaN scenario, 
in Ref.~\cite{Bjaelde:2007ki} Bjaelde $et~al.$ have argued that a negative value of the sound speed does 
$\it not$ always indicate the occurance of an instability. 
Due to the dragging force of cold dark matter, 
neutrino perturbations can remain stable even if the sound speed of the dark energy fluid becomes negative.
More specifically, the dragging force due to the dark matter and baryons can 
stabilise the neutrino perturbations if the condition \cite{Bjaelde:2007ki},
\begin{equation}
\left(\frac{\Omega_{CDM} + \Omega_{b}}{\Omega_{\nu}}\right)
\left(\frac{G}{G_{eff}}\right)
\left(\frac{\delta_{CDM}}{\delta_{\nu}}\right) \gg 1 \label{compare}
\end{equation}
is satisfied, where
\begin{equation}
G_{eff} = G\left(1 + \frac{2 f^2(A) M_{Pl}^2}{1 + \frac{a^2}{k^2}(V''(A) +
\rho_{\nu} f'(A)  )   }\right), \label{newton}
\end{equation}
and where the coupling function $f(A)$ is defined as 
\begin{eqnarray}
f(A) = \frac{1}{\bar m_{\nu}} 
\frac{d \bar m_{\nu}}{dA}\;,
\end{eqnarray} 
with $f(A) =-\frac{1}{A}$ for the case $\beta=1$.
Using the above definitions the stability condition of Eq.~(\ref{compare}) can
be rewritten as 
\begin{equation}
f^2(A)< \left(\frac{\Omega_{CDM}  + \Omega_{b}  - \Omega_{\nu}}   {2
\Omega_{\nu} M_{Pl}^2}\right)
\left(\frac{\delta_{CDM}}{\delta_{\nu}}\right). \label{stabilitycondition}
\end{equation}
From this equation we can see that a small value of $\Omega_{\nu}$ (i.e.\ a small 
neutrino mass scale today) as well as a larger value of $\delta_{CDM}$ 
compared to $\delta_{\nu}$ can help to stabilise the neutrino perturbations.

Setting ${\delta_{CDM}}/{\delta_{\nu}}\sim 1$ for simplicity and using Eq.~({\ref{stabilitycondition}}), we find that, e.g., for $\bar m_{\nu}^{(0)} \sim 1.5$ eV, $f(A)$ has to be smaller 
than $2 M_{Pl}^{-1}$ or for $\bar m_{\nu}^{(0)} \sim 0.06$ eV, $f(A)$ has to be 
less than $10 M_{Pl}^{-1}$. 
In order to avoid the instability the 
present value of $A$ should therefore be larger than about 
$M_{Pl}/10$ for $\bar m_{\nu}^{(0)} \sim 0.06$ eV
and $M_{Pl}/2$ for $\bar m_{\nu}^{(0)} \sim 1.5$ eV.
For the case $\bar m_{\nu}^{(0)} \sim 0.06$ eV this condition is well compatible with the 
adiabaticity condition in Eq.~(\ref{kinoverdensity}) and the simple analytical consideration suggests that
the instability problem might be cured. 
On the other hand, for $\bar m_{\nu}^{(0)} \sim 1.5 ~eV$ the adiabaticity and the stability conditions 
cannot safely be satisfied simultaneously and we conclude that whether the instability appears has to be checked numerically.\footnote{We note that the calculation of sound speed (see e.g.\ \cite{Takahashi:2006jt}) does not depend on the present value of $A$ and it is therefore still negative in our model, however, as we have mentioned before we are following the argument of \cite{Bjaelde:2007ki} that despite a negative sound speed the instability can be avoided by the dragging force of dark matter.}

Finally, we would like to note that, as we have discussed in section 4, for the most standard types of  
potentials (e.g.\ with $\alpha=2,\beta=1$) the dynamical 
contribution to the total dark energy is rather small (typically a few percent), 
and the main contribution arises from a constant vacuum energy density.
This means that even if the neutrinos would cluster, 
the observable effects on the smoothness of dark energy are much smaller than in the 
standard MaVaN scenario.
Another related question is whether the scales on which neutrinos might cluster finally leads to any observable effect \cite{Mota:2008nj}. We conclude by remarking that although we have given some arguments 
which suggest that stability problems for 
non-relativistic neutrinos could be resolved in certain MaVaN Hybrid Scenarios with 
``high-scale'' seesaw, this issue requires further investigations (which are left for future studies).

\section{Summary and Conclusions}
Motivated by the intriguing proximity of the energy density of dark energy and the neutrino mass scale 
we have studied the phenomenology of hybrid scenarios of neutrino dark energy, where in addition to a so-called Mass Varying Neutrino (MaVaN) sector, a cosmological constant (from a false vacuum) is driving the accelerated expansion of the universe today. 
Within the generalised framework we have focused on phenomenological issues such as on the connection to the neutrino mass scale and on its consequences for the dynamical nature of dark energy.  

We have therefore calculated the effective equation of state parameter $w_{eff}(z)$ in the MaVaN Hybrid Scenario where the effective potential for the dynamical real scalar field (the ``acceleron'' field $A$)
has the following form
\begin{equation}
V(A)_{eff} =  M^{4- \alpha}A^{\alpha} + \frac{\rho_{\nu}^{(0)}}{a^3} \left(\frac{A_0}{ A}\right)^{\beta} +  V_{0} \;. 
\end{equation}
We found that, for the case of a power law potential in the MaVaN sector, $w_{eff}(z)$ is determined by the neutrino mass scale and by the parameters $\alpha$ and $\beta$ (c.f.\ Eq.~(\ref{Eq:weff_mnu})). Due to the interactions of the dark energy field with the neutrino sector, $w_{eff}(z)$ is predicted to become smaller than $-1$ for increasing $z>0$ (c.f.\ Fig.~\ref{weffective_plot}), which could be tested in future cosmological observations.  
For the considered scenarios, we have also calculated how the neutrino mass scale determines which fraction of the dark energy is dynamical, and which originates from the ``cosmological constant like'' vacuum energy. 
In particular, for the case of a mass term potential and a standard seesaw relation (i.e.\ $\alpha=2$ and $\beta=1$) we found that compatibility with the terrestrial neutrino mass bounds requires a large contribution of constant vacuum energy (c.f.\ Eq.~(\ref{Eq:MnuvsGamma})).

Another interesting question, which we have investigated in the MaVaN Hybrid Scenario with power law potentials is whether it is possible to realise neutrino dark energy with a ``high-scale'' seesaw mechanism, where the right-handed neutrino masses are close to the GUT scale. We found that the field value of the ``acceleron'' field as well as the masses of the right-handed neutrinos can indeed be large and in principle a hybrid scenario of neutrino dark energy might be realised with a ``high-scale'' seesaw. We have also commented on how the Hybrid MaVaN Scenarios with  ``high-scale'' seesaw might allow to suppress the formation of ``neutrino nuggets'' and resolve stability problems of dark energy models with non-relativistic neutrinos.  

In summary, we have found that the considered hybrid scenarios of neutrino dark energy have several attractive features, in particular the close connection to the neutrino mass scale and an effective dark energy equation of state parameter $w_{eff}(z)$ which depends only on the parameters $\alpha$ and $\beta$ of the potential and on the neutrino mass scale. The prediction that $w_{eff}(z) < -1$ for $z>0$ provides a ``smoking gun'' signal for such interacting dark energy scenarios, which could be observed in future surveys. Issues which are still open, and which are left for further studies, are whether a MaVaN Hybrid Scenario with ``high-scale'' seesaw can indeed solve the stability problems of conventional MaVaN scenarios with non-relativistic neutrinos and whether a consistent model can be constructed where this is realised.

\section{Acknowledgements}
We would like to thank Neal Weiner for useful discussions. This work was partially supported by 
The Cluster of Excellence for Fundamental Physics 
``Origin and Structure of the Universe'' (Garching and Munich). SD is supported by US NSF CAREER 
grant  PHY-0449818 and  DOE OJI program under grant  DE-FG02-06ER41417. SD acknowledges the hospitality of Max-Planck-Institut f\"ur Physik (Werner-Heisenberg-Institut) in Munich where this work was initiated.

\section*{Appendix}
\appendix

\renewcommand{\thesection}{\Alph{section}}
\renewcommand{\thesubsection}{\Alph{section}.\arabic{subsection}}
\def\theequation{\Alph{section}.\arabic{equation}}
\renewcommand{\thetable}{\arabic{table}}
\renewcommand{\thefigure}{\arabic{figure}}
\setcounter{section}{0}
\setcounter{equation}{0}

\section{Some Remarks on the possible Origin of the MaVaN Hybrid Potential}

\subsection{Fractional Power Potential from Non-Canonical Kinetic Energy Terms}
One possible origin of fractional powers in the potential in Eq.~(\ref{potential}) 
is a non-canonically normalised real scalar field $A'$ with the
following Lagrangian
\begin{equation}
{\mathcal L} = \frac{A'{}^{n'}}{M_{1}^{n'}} \left(\partial_{\mu}A' \right)^2
- V(A')\:,
\end{equation}
where $V(A') =  M_{2}^{4-\alpha '} A'{}^{\alpha '}$. 

In the MaVaN Hybrid Scenario, the effective energy density of the field $A'$
has the form 
\begin{equation}
V_{eff}(A') =  \frac{A'{}^{n'}}{M_{1}^{n'}} \left(\partial_{\mu}A' \right)^2
+ M_{2}^{4-\alpha '} A'{}^{\alpha '} + \frac{\rho_{\nu}^{(0)}}{a^3} \left(
\frac{A'_0}{A'}  \right)^{\beta '}\,,
\end{equation}
where the last term arises form the smooth background energy density. We
have assumed $\bar m_{\nu}(A') \sim \frac{1}{{A'}^{\beta'}}$. 

Now we would like
to write it in terms of canonically normalised field $A$, which satisfies 
\begin{equation}
\frac{1}{\sqrt{2}}\partial_{\mu}A = \left(\frac{A'}{M_{1}} \right)^{n'/2}\partial_{\mu}A'\;.
\end{equation}
Integrating this equation we obtain the canonically normalised field $A$ in
terms $A'$,
and the effective potential in terms of the normalised field $A$
becomes,
\begin{equation}
V_{eff}(A) = g^{\alpha '} M_{2}^{(4-\alpha ')}M_{1}^{\frac{n'\alpha
'}{n'+2}}A^{\frac{2\alpha'}{n'+2}} + \frac{\rho_{\nu}^{(0)}}{a^3}
\left(\frac{A_0}{A}\right)^{\frac{2\beta'}{n'+2}},
\end{equation}
where $ g = \left(n'+2\right)^{\frac{2}{n'+2}}/2^{\frac{3}{n'+2}}$ which can be easily
absorbed in the redefinition of $M$. 

As a consequence of the non-canonically normalised scalar field we have obtained
fractional powers in the MaVaN Hybrid potential with $\alpha = \frac{2 \alpha '}{n' + 2}$
and $\beta = \frac{2 \beta '}{n' + 2}$.

\subsection{Generalisation to Three Families: Possible Origin of $V_0$}
One motivation for introducing a constant term $V_0$ to the potential arises
in supersymmetric ``hybrid-type'' models of dark energy. Generalising 
the superpotential of Eq.~(\ref{Eq:Origin_1}) to three families, we may 
realise the situation that some of the ``waterfall fields'' $N_i$ 
are still in the false vacuum while others are in the true
vacuum where $|N_i| \sim v_{N_i}$. A simple superpotential with this
characteristic feature is the following
\begin{eqnarray}\label{Eq:threefamilies}
W = \lambda \hat A (x_i \hat N_i^2 - v_{N_i}^2) + y_{\alpha i} \hat H^0_u \hat
\nu_\alpha  \hat N_i \;.
\end{eqnarray}
In this case, the vacuum energy of the acceleron field does not account for
the total dark energy, but only for part of it. The other part stems from the $N_i$
which are in the false vacuum leading each to a contribution of energy
density $\lambda^2  v_{N_i}^4$. It is also interesting to note that the $N_i$ which 
have non-zero vevs generate a mass term for $A$ from the contribution 
$|F_N|^2$ to the scalar potential.

\subsection{Higher Powers of A from the Superpotential}
It is furthermore possible to generalise the setup of Eq.~(\ref{Eq:threefamilies}) 
such that higher powers of $A$ arise in the potential. Such higher powers can 
emerge from a superpotential of the form
\begin{eqnarray}
W = \lambda \frac{A^n}{M^{n-1}} (x_i \hat N_i^2 - v_{N_i}^2) + y_{\alpha i}
\hat H^0_u \hat \nu_\alpha  \hat N_i \;.
\end{eqnarray}
The most relevant parts of the scalar potential (for a $N_i$ in the true
vacuum) would then be given by
$|F_{A}|^2 \sim v_{N_i}^4 |A|^{2n-2}$ and $|F_{N_i}|^2 \sim |\frac{\lambda
A^n}{M^{n-1}}N_i + y_{\alpha i} \hat H^0_u \hat \nu_\alpha|^2$.
This generalisation of the superpotential has two effects: Firstly,
$|F_{A}|^2$ leads to a higher power of A, explicitly $|A|^{2n-2}$, i.e.\
$\alpha=2n-2$ in the effective potential of Eq.~(\ref{effective_potential}) 
and secondly, $|F_{N_i}|^2$ results in masses of the right-handed neutrinos
of the form $m_{N_i}^2 \sim \frac{x_i \lambda^2 |A|^{2n}}{M^{2n-2}}$ which
leads to masses of the light neutrinos $\bar m_{\nu} \sim
\frac{M^{n-1}}{|A|^n}v^2$. This implies $\beta=n$ in Eq.~(\ref{effective_potential}).
We note that another way to realise higher powers of $\beta$ would be to
generate neutrino masses via a multiple seesaw (see e.g.\
\cite{Bhatt:2007ah} for a model of neutrino dark energy using the double seesaw
mechanism).

\end{document}